\documentstyle[prb,aps,eqsecnum,manuscript]{revtex}

\begin{document}

\title{Temperature effects on dislocation core energies \\ in silicon and 
germanium}
 
\author{Caetano R. Miranda}

\address{Instituto de F\'{\i}sica ``Gleb Wataghin'', 
Universidade Estadual de Campinas,\\
CP 6165, CEP 13083-970, Campinas, SP, Brazil}       

\author{R. W. Nunes }

\address{Departamento de F\'{\i}sica, Universidade Federal de Minas Gerais, \\
Belo Horizonte, Minas Gerais , CEP 30123-970, Brazil}       

\author{A. Antonelli}

\address{Instituto de F\'{\i}sica ``Gleb Wataghin'', 
Universidade Estadual de Campinas,\\
CP 6165, CEP 13083-970, Campinas, SP, Brazil} 

\date{\today}
\maketitle
   
\begin{abstract}

Temperature effects on the energetics of the 90$^\circ$ partial
dislocation in silicon and germanium are investigated, using
non-equilibrium methods to estimate free energies, coupled with Monte
Carlo simulations. Atomic interactions are described by Tersoff and
EDIP interatomic potentials. Our results indicate that the vibrational
entropy has the effect of increasing the difference in free energy
between the two possible reconstructions of the 90$^\circ$ partial,
namely, the single-period and the double-period geometries. This effect
further increases the energetic stability of the double-period
reconstruction at high temperatures. The results also indicate that
anharmonic effects may play an important role in determining the
structural properties of these defects in the high-temperature regime.

\end{abstract}

%\maketitle

\pacs{61.72.Lk, 61.72.Bb, 62.20.Fe}

%\begin{multicols}{2} 

%
\section{Introduction}              

The study of the atomic structure of dislocation cores in
semiconductors is not free from outstanding
issues.~\cite{hirsch,duesbery,alexan,bulatov} At low deformations,
dislocations in these systems are known to occur along $<$110$>$
directions on $\{111\}$ slip planes of the diamond (or zinc-blend)
lattice. Due to the presence of a two-atom basis, this lattice
supports two possible $\{111\}$ dislocation slip planes, the so-called
glide and shuffle sets.  In the glide set, the slip plane lies between
atomic layers separated by a third of the nearest-neighbor bond
length, and geometrically it would appear that dislocation glide
should require breaking three covalent bonds.  In the case of the
shuffle set, the slip plane lies between atomic layers separated by a
full bond length, and glide appears to involve breaking only a single
covalent bond. This would suggest that lattice friction should be
smaller in the shuffle set, and hence that dislocations would glide in
that configuration. However, the prevalent opinion is in favor of the
glide set, based primarily on the experimental evidence showing that
dislocations in Si are dissociated into Shockley partials bounding
stacking-fault ribbons, and on the fact that stacking faults are not
stable in the shuffle plane.~\cite{hirsch,duesbery,alexan,bulatov}
Nevertheless, this conclusion has been disputed by Louchet and
Thibault-Desseaux,~\cite{louchet} who argued that partials in the
shuffle position could occur in conjunction with glide-plane stacking
faults.  The mechanism controlling dislocation glide is also a matter
of debate. Many theoretical and experimental works have either
supported or assumed the Hirth and Lothe mechanism,~\cite{hl} based on kink
nucleation and migration, but an impurity-pinning
mechanism proposed by Celli et al.~\cite{celli} has been shown to be
consistent with recent experimental results.~\cite{kolar} This issue
is far from being resolved.

Another unresolved issue has recently arisen, regarding the core
structure of the 90$^\circ$ partial dislocation. In $\{111\}$ planes,
dislocations lining along $<$110$>$ directions have their Burgers
vectors either parallel (the screw orientation) or making a 60$^\circ$
angle with the dislocation line. Both dissociate into Shockley
partials, the former into two 30$^\circ$ partials, and the latter into
a 30$^\circ$ partial and a 90$^\circ$ partial.  Both partials are
believed to undergo core reconstruction, as indicated by the low
density of unpaired spins found in EPR
experiments.~\cite{hirsch,duesbery,alexan,bulatov}

In the glide set, the nature of the reconstruction of the 90$^\circ$
partial appears straightforward. The symmetrically-reconstructed
geometry shown in Fig.~\ref{cores}(a) displays mirror symmetries along the
$[110]$ direction. A symmetry-breaking reconstruction shown in
Fig.~\ref{cores}(b) lowers the core energy by $\sim$200 meV/\AA, according to
{\it ab initio} and tight-binding (TB)
calculations.~\cite{Bigger,nbv,hansen} This structure retains the
periodicity of the lattice along the dislocation direction, thus being
called the single-period (SP) structure. Until recently, it had been a
matter of consensus that the SP core was the ``ground-state''
structure of the 90$^\circ$ partial, since the earliest independent
works of Hirsch and Jones.~\cite{hirsch2,jones} However, the structure
shown in Fig.~\ref{cores}(c) has been recently proposed by Bennetto and
collaborators,~\cite{Nunes_2} and shown to have lower internal energy
than the SP core, by means of total-energy tight-binding (TETB), and
{\it ab initio} local-density-approximation (LDA) calculations. In
this new geometry, along with the mirror symmetry, the translational
symmetry of the lattice by a single period is also broken, and the
periodicity along the core is doubled. For that reason, this
reconstruction has been called the double-period (DP) structure.

Both the SP and DP geometries are consistent with all available
experimental information about the 90$^\circ$ partial. EPR
measurements in Si indicate a rather small density of dangling bonds
in the core of the dislocation.~\cite{hirsch,duesbery,alexan} The two
structures are fully reconstructed, meaning that neither would give
rise to deep-gap states which would show an EPR signal. Moreover, both
cores consist entirely of fivefold, sixfold, and sevenfold rings, both
being consistent with images produced by transmission electron
microscopy, at the current level of resolution of this
technique.~\cite{kolar} So far, the experiments appear unable to
decide clearly on the issue. Recent experimental work by
Batson~\cite{batson} indicates a DP-derived structure (called the
``Extended DP structure'' by this author) to be more consistent with
STEM and EELS experiments.

At the theoretical level, several calculations have addressed the
energetics of the SP and DP cores in diamond (C), silicon (Si), and
germanium (Ge), using different
methods.~\cite{Nunes_2,Vall,nv,Blase,Nunes_1,Nunes_3,Lehto} Virtually
all of the more accurate {\it ab initio} and TETB
studies~\cite{Nunes_2,nv,Blase,Nunes_1,Nunes_3} agree that the DP core
is more stable at 0 K, with the exception of the {\it ab initio}
cluster calculations for Si in Ref.~\onlinecite{Lehto}. For example,
LDA and TETB calculations in Refs.~\onlinecite{Nunes_2} and
\onlinecite{Nunes_3} indicate that the internal energy of the DP
structure is lower than that of the SP core in all three materials (C,
Si, and Ge). A possible influence of supercell boundary conditions on
these results was raised in Ref.~\onlinecite{Lehto} (using
Keating-potential calculations) and later refuted by supercell-size
converged TETB calculations in Ref.~\onlinecite{nv}. The issue was
further revisited in Ref.~\onlinecite{Blase}, with {\it ab initio}
calculations showing that the DP core remains more stable in C even
under relatively severe stress regimes.  Ideally, supercell-size
converged calculations like the TETB ones in Ref.~\onlinecite{nv}
would properly address the issue for isolated dislocations in the
bulk. However, in that study very large supercells (containing up to
1920 atoms) were employed, which makes such calculations prohibitive
for the more accurate {\it ab initio} techniques.

Closely related to the present study is the work of Valladares and
co-workers,~\cite{Vall} in which temperature effects on the energetics
of the SP and DP cores in Si were investigated using a Tersoff
potential,~\cite{Ters} within a harmonic approximation treatment of
vibrational entropic effects in the dislocation free energy, at finite
temperatures. They find that the free-energy difference between the
two cores, at a temperature of 800 K, becomes smaller than the
zero-temperature internal energy split, suggesting the possibility of
the two cores playing a role in the plasticity of Si, in the
temperature range of most plastic deformation experiments ($\sim$700-1000
K). However, the calculations in Ref.~\onlinecite{Vall} did not
include anharmonic effects, which are likely to be relevant in that
range of temperatures.

In this paper, we present the results of a study of anharmonic
vibrational effects in the free-energy difference between the SP and
DP cores in Si and Ge. As discussed in the following, we arrive at
different conclusions with respect to the work of Valladares {\it et
al.}, with our results indicating that the effect of increasing
temperature is to enhance the thermodynamical stability of the DP core
with respect to the SP geometry.  In our calculations, anharmonic
effects are fully taken into account by using a methodology that
allows to estimate the free energy through non-equilibrium
simulations.  The details of this methodology are discussed in
Sec. II. In Sec. III our results are presented and discussed. We end
with a summary and conclusions in Sec. IV.

\section{Methodology}              

\subsection{Interatomic Potentials}
\label{secA}

There are several empirical potentials available in the literature to
model the interactions between Si atoms.~\cite{Potential1} In the
present work, our aim is to account for vibrational entropic effects
on the energetics of the SP and DP reconstructions of the 90$^{\circ}$
partial dislocation in Si. One key ingredient for this purpose is that
both the SP and DP structures be a local minimum of the potential. It
is essential then that the potential be able to describe the
reconstruction-driven symmetry breaking that leads to the SP
geometry.~\cite{Bigger,nbv,hansen} This is an important aspect, since
both core structures consist of fully reconstructed bonds, and a
realistic description of vibrational properties requires a proper
treatment of bonding for the minima in question. For example,
potentials such as the Stillinger-Weber(SW) and the Kaxiras-Pandey(KP)
fail in that regard.~\cite{Potential2} Both the Tersoff~\cite{Ters}
and the Environment Dependent Interatomic Potential
(EDIP)~\cite{edip1,edip2} models describe correctly the energetics of
the SP reconstruction with respect to a symmetric
(so-called quasi-fivefold) structure.~\cite{Bigger,edip2}

The focus of our paper is on the effects of the dislocation
vibrational modes on the free energies of the two core models at
finite temperatures.  In a previous work we have used these two
potentials to estimate the free energy of Si in several different
phases (amorphous, liquid, and crystalline, as well as clathrate
structures), obtaining results which are in good agreement with those
of experimental and other theoretical works.~\cite{Caetano} For a more
specific analysis of the description of anharmonic effects by the EDIP
and Tersoff potentials, we show in Fig.~\ref{bulk} the results we obtained for
the free energy as a function of temperature, in a 216-atom bulk
supercell.  The figure includes also the Tersoff-potential harmonic
approximation values from Ref.~\onlinecite{harm} and the experimental
results from Ref.~\onlinecite{fexp}. As expected, the experimental
numbers start deviating appreciably from the harmonic approximation
curve around the Debye temperature of Si ($\sim$640 K), where
anharmonic effects begin manifesting themselves. From the figure, we
see that both the Tersoff and the EDIP potentials can account for
these anharmonic effects in the free-energy curve at a quantitative
level.

Moreover, it is clear that the EDIP results are in better agreement
with the experimental ones. Of particular relevance to the present
work is the scale of the difference in free energy between
harmonic-approximation and experimental results, which is
$\sim$10~meV/atom at 640 K. As a first approximation, we may consider
ten atoms per period as constituting the core in the SP and DP
structures (those taking part in the fivefold and sevenfold rings,
seen on [110] projections, surrounding the geometric center of the
core), which gives an initial estimate of $\sim$30~meV/\AA\ for the
scale of the anharmonic effects, from the bulk results. From this, one
might expect that anharmonic effects should be important in the case
of the 90$^\circ$ partial, since the difference in enthalpy at 0 K
between the SP and DP cores is $\sim$60~meV/\AA.

On the basis of these results, we argue that both potentials give a
realistic description of vibrational effects, in particular of the
anharmonic contributions which become important in the range of
temperatures of our study. For the Ge case, only a Tersoff model was
employed, since there are no available parameters for Ge in the EDIP
form. The functional form and the parameters we use for the two models
are given in Refs.~\onlinecite{Ters} (Tersoff) and \onlinecite{edip2}
(EDIP).

\subsection{Simulations and Free Energy Calculations}
\label{secB}

In our statistical computation of dislocation free energies, we use a
Metropolis-algorithm Monte-Carlo method. Two types of simulations were
performed, one employed a canonical-ensemble (constant-NVT) treatment,
and the other simulated an isobaric-isothermal ensemble
(constant-NPT).~\cite{Frenkel} A simulation cell with 192 atoms was
used, containing a 90$^\circ$ partial-dislocation dipole which was
introduced by displacing the atoms according to the
continuous-elasticity solution for the displacement field of a
dislocation dipole.~\cite{hl} The introduction of a dipole enables the
use of periodic boundary conditions to avoid surface effects.  It
should be pointed out that the free energy calculations presented in
this work are far more demanding in terms of computer time than
harmonic-approximation calculations, which only require one
diagonalization of the dynamical matrix, in order to determine the
phonon frequencies. Therefore, the complexity of our calculations
imposes a limitation on the size of the cell used in the
simulations. The cell size chosen reflects the best compromise between
a good description of the essential physics and computational
feasibility.

The free energy calculations were performed using a non-equilibrium
method, namely the Reversible-Scaling method (RS).~\cite{RS_1} This
method consists of simulating a quasi-static process where the
potential-energy function of the system is dynamically scaled. It
allows for an accurate and efficient evaluation of the free energy
over a wide range of temperatures, using a single Monte Carlo
simulation, starting from a reference temperature for which the free
energy of the system is known. In order to obtain the free energy at
this reference temperature, we used the Adiabatic Switching method
(AS).~\cite{AS_1} Using the AS approach, the absolute free energy can
be computed by determining the work done to switch the potential
energy of the system from the configuration of interest to that of a
reference system for which the free energy is known analytically. The
systematic error in the free energy, caused by dissipation, can be
estimated by reversing the switching procedure and computing the
overall work performed in this round-trip procedure. The statistical
error can be estimated by performing several trajectories in the
configuration space of the system.

In this work, the reference system was an Einstein crystal, i.e., a
collection of independent harmonic oscillators. The angular frequency
of the oscillators ($\omega$) was taken to be approximately the
vibration modes of Si ($\omega = 28$~Trad/s) and Ge ($\omega = 15$~
Trad/s), which correspond to the acoustic phonon modes from the
edge of the Brillouin zone that give rise to a pronounced peak in the 
phonon density of states of these materials.~\cite{Kane,Nilsson}
It should be pointed out that the final value for the Gibbs free energy
is not sensitive to the choice of the oscillator frequency,
provided that the chosen frequency is not very far from the
relevant frequencies of the physical system.
Monte Carlo constant-NVT simulations were used to generate
switching trajectories, with a typical "switching time" of 3.0
$\times$ 10$^{5}$ MC/steps. The volume in these simulations was chosen
to be that obtained from a constant-NPT simulation for the given
reference temperature at zero pressure. In general, 25 switching
trajectories were used to estimate the statistical error. The
estimated error was $\pm$ 1.0 $\times$ 10$^{-3}$ eV/atom. RS
calculations were carried out at zero pressure, covering a range from
T=100 K to T=2000 K, using a scaling time of 5.0 $\times$ 10$^{5}$
MC/steps. Five different switching trajectories were sampled, in order
to estimate the statistical errors. In this way, we were able to
estimate the absolute Gibbs free energy (G$_{abs}$) at zero pressure
of the SP and the DP reconstructions in the interval from 300K to 1200
K using EDIP and Tersoff models for Si, and for Ge in the same range
of temperatures using a Tersoff potential. The absolute vibrational
entropy was calculated by taking the numerical derivative of the
absolute Gibbs free energy with respect to temperature: $S_{abs}=
-(\frac{\partial G_{abs}}{\partial T})$.

\section{Results and Discussion}              
\label{enthal}
\subsection{Enthalpies at zero Kelvin}

Imposing periodic boundary conditions in a dislocation calculation
using supercells requires the net Burgers vector in the cell to
vanish. This is done by using cells which contain two dislocations
with opposite Burgers vectors. Hence, besides the core enthalpies that
interest us, supercell calculations also include elastic interactions
between the two dislocations, and between the dislocations and their
periodic images. In order to extract meaningful core energy values,
these elastic interactions must be properly accounted for. Previous
calculations have shown that dislocation cores in covalent
semiconductors are very narrow, with the elastic fields reaching the
linear behavior predicted by the continuum linear elasticity solution
already at a distance of about two lattice periods from the center of
the core.~\cite{hansen,Blase} This indicates that, to the extent that
one is interested in differences in energies between different core
models, long-range elastic effects should mostly cancel out for
sufficiently large cells. This is because these long-range fields are
defined by the dislocation slip system which is obviously the same in
our case, since we are comparing different core structures of the same
dislocation. We must be careful, however, in providing numerical
evidence that the supercells we use in our calculations are
sufficiently large, in the above sense.

In the particular case of the SP geometry, it is known that a
relatively strong dipolar elastic interaction is present, which
depends on the relative senses of reconstruction of the two
dislocations in the cell, with respect to the broken mirror
symmetry. Depending on the size of the cell, this interaction may be
of the same order of the core-energy difference between the SP and DP
geometries. (A similar effect is present in the DP core, but this is a
weaker quadrupolar interaction which can be neglected.~\cite{Nunes_2}) As
discussed in the literature,
~\cite{Nunes_2,Vall,nv,Blase,Nunes_1,Nunes_3} this dipolar interaction
can be properly handled by considering the SP-core energy as the
average energy between two supercell calculations, one in which the
two SP dislocations are reconstructed in the same sense, and the other
where they have opposite reconstruction
senses.~\cite{nv,Nunes_1,Nunes_3} In order to further investigate the
accuracy of this procedure and also to address the numerical
convergence level of our calculations for the cell sizes we use, we
show in Table I the 0 K enthalpies computed with the Tersoff and EDIP
potentials for several different cell sizes. For comparison, we also
include the TETB results from Ref.~\onlinecite{nv}, to which we add
numbers for a 240-atom supercell. In the notation introduced in
Ref.~\onlinecite{Lehto}, the relevant supercell parameters are $D$,
the height of the cell in the $<111>$ direction (perpendicular to the
slip plane), and $L$ the width of the cell in the $<11\bar{2}>$
direction (perpendicular to the dislocation line) on the slip
plane. We consider here only supercells with ``dipolar'' boundary
conditions,~\cite{bc} which were shown by Lehto and \"Oberg~\cite{Lehto} to
minimize the strains associated with the stacking of the infinite
array of dislocations. We will comment below on the comparison of our
results with those from Ref.~\onlinecite{Vall}, which were obtained
with ``quadrupolar'' boundary conditions.~\cite{bc}

Table I shows our results for $E_{SP-DP} = \overline{E}_{SP} -
E_{DP}$, the energy difference between the average energy of the two
SP cells ($\overline{E}_{SP}$) and the energy of the DP cell, and for
$\Delta E_{SP} = E_{SP}^{same} - E_{SP}^{opposite}$, the energy
splitting between the two SP cells (as described in the previous
paragraph), for four different cell sizes. Note that $E_{SP-DP}$
converges for relatively small cell sizes, and faster than $\Delta
E_{SP}$, due to the cancellation of elastic effects that happens when
we take the average of the two SP calculations. For the free-energy
calculations we discuss in the next section, we used the smaller cell,
with $L = 13.3$~\AA\ and $D = 18.8$~\AA, for a total of 192 atoms. In
the 240 atom cell, $D$ is the same as in the 192-atom cell, and $L$ is
only 25\% larger.  We can see from Table I that, already for this
240-atom cell, the TETB result for $E_{SP-DP}$ is converged to within
the numerical accuracy of the method. This is also what can be
observed for the EDIP potential. The EDIP result for $E_{SP-DP}$ in
the 192-atom cell is within 13\% of the EDIP converged value, and for
the 240-atom cell $E_{SP-DP}$ is converged to within 3\%.  The Tersoff
numbers converge more slowly, with the 192-atom cell numbers being
more than twice as large as the converged value.  In the case of Ge,
we only calculated $E_{SP-DP}$ using the 192-atom cell. We obtained
$E_{SP-DP}$~= -8~meV/\AA\ in disagreement with the LDA and
Keating-potential results from Ref.~\onlinecite{Nunes_1}. However, as
discussed in Sec.~\ref{free-ener}, the behavior of the free energy as
a function of temperature for Ge is the same as that for Si, i.e.,
$E_{SP-DP}$ increases as the temperature is raised.  If we were to
extrapolate to the free-energy calculations these convergence trends
observed for the 0 K enthalpy results in Table I, we would expect our
EDIP free-energy values to be converged to within $\sim$10\%, and the
Tersoff values to be about twice as large as what we would obtain with
larger cells.

The TETB calculations in Table I were done at the fixed cell volume
(corresponding to the experimental lattice constant). As a test, we
also performed calculations for volumes 1\% larger and smaller,
obtaining the same convergence trends, i.e., energies were essentially
converged for the 240-atom cell. More importantly, the convergence
calculations with the Tersoff and EDIP potentials in Table I were done
at constant pressure, resulting in very small changes from the initial
volumes.

Before proceeding to the discussion of the free-energy results, let us
put the results in Table I in perspective. Previous studies have shown
that the Tersoff potential does give the right ordering of 0 K
enthalpies, despite underestimating the difference in enthalpies
between the two cores. Our converged Tersoff values of 6~meV/\AA\ are
in very good agreement with the converged results obtained in
Ref.~\onlinecite{chrzan} using supercell and cylindrical cluster
calculations. However, they are smaller by one order of magnitude,
when compared with the converged TETB values. The EDIP potential has
the shortcoming of predicting the wrong ordering of 0 K enthalpies for
the SP and DP cores. When comparing our results with those obtained by
Valladares {\it et al.}, we must observe that they used cells with $D
= 9.4$~\AA, which is half the height of our cells, and quadrupole
boundary conditions. As a check, we did run Tersoff calculations with
the same cell parameters and obtained the same results as in their
work. The value for $E_{SP-DP}$ in this case is 32 meV/\AA, which is
more than five times as large as the converged Tersoff value in Table
I.  This shows that, at least for the 0 K enthalpy, their results are
not converged at the same level as our calculations.

\subsection{Free energies}
\label{free-ener}

In order to test whether the empirical potentials support the SP and
DP reconstructions at high temperatures, we performed constant-NPT
simulations of all structures in a broad range of temperatures, from 0
K up to 2000 K (1500 K) for Si (Ge). The structures were further
characterized for several temperatures (at every 500 K between 0 K and
the highest temperature) by evaluating structural properties such as
the pair-correlation function, the bond-angle distribution, and the
atomic-coordination number. Only the natural thermal broadening of
these functions was observed, with no structural changes and all atoms
remaining fourfold coordinated. From these results, we conclude that
for both the Tersoff and the EDIP potentials, the SP and DP
reconstructions remain stable and do not undergo restructuring over
the entire range of temperatures we tested for each of the two
semiconductors. 

The metastability of the SP core, even at temperatures as high as 2000
K, is not surprising. Unlike the spontaneous transformation of the
symmetric quasi-fivefold core into the SP core, which happens at 0 K
due to the absence of an energy barrier,~\cite{Bigger,nbv,hansen} the
SP $\rightarrow$ DP transformation involves breaking the strong
reconstructed covalent bonds in the SP core. The same bond breaking
mechanism regulates the kink migration barriers.~\cite{nbv} We have
computed an energy barrier of $\sim$1.5 eV for the process involving
the conversion of a segment of two SP periods into a segment of one DP
double period.
    
Let us now turn to the focus of this work, the Gibbs free-energy
difference, $\Delta G$, between the SP and the DP cores as a function
of temperature, obtained by the Reversible Scaling Method within the
Monte Carlo method (RS-MC). In the previous section we discussed the
numerical convergence of our calculations with respect to the two
dimensions of the supercell which are perpendicular to the dislocation
line. When we introduce thermal effects, periodicity along the line is
disrupted by atomic vibrations, and one must be careful about the
sampling of the phonon modes along the dislocation direction. In our
free-energy calculations, we used 192-atom supercells with twice the
lattice period along this direction. To test the phonon sampling in
this cell, we also calculated free energies for a supercell with four
times the lattice period along the line, using the Tersoff
potential. In Fig.~\ref{tamanho} we show the results for the
free-energy difference $\Delta G$ between the SP and DP cores, for the
two types of SP cells and also for the average between the two
cells. Note that, while the results for $\Delta G$ for the individual
SP calculations differ quite appreciably for a given cell size and
vary substantially when we go from the smaller (192 atoms and twice
the lattice period along the dislocation line) to the larger cell (384
atoms and four times the period along the line), the results for
$\Delta G$ between the average of the SP cells and the DP cell are
very similar for the two cell sizes. This shows that our simulations
with 192 atoms capture most of the relevant differences between the
vibrational modes of the two cores and also lends further confirmation
of the cancellation of elastic interaction effects that occurs when
the average between the two SP cells is considered, as discussed in
the previous section.

In Fig.~\ref{G-si} we show the results for $\Delta G_{SP-DP}$ for the
Tersoff and EDIP models in a 192-atom supercell, showing now only the
average of the SP calculations. Fig.~\ref{G-ge} shows the results for
Ge using the Tersoff potential. In order to analyze the entropic
effects, Figs.~\ref{G-si} and \ref{G-ge} also display the entropic
contribution ($-T \Delta S$) to $\Delta G$. The Tersoff results for Si
in Fig.~\ref{G-si} show $\Delta G$ increasing with temperature from
19~meV/\AA\ at 300 K to 37~meV/\AA\ at 1200 K.  The entropic term also
increases with temperature, being the dominant contribution for
$\Delta G$ in the high temperature regime.  For Si using the EDIP
model, we observe the same behavior, with $\Delta G$ increasing with
temperature from -32~meV/\AA\ at 300 K to -23~meV/\AA\ at 1200 K,
notwithstanding the fact that for this potential the SP core remains
more stable over the entire temperature range. This is mostly due to
the fact that the EDIP potential does not describe properly the
energetics of the two reconstructions at 0 K, as discussed above. But
even for this potential, the behavior of the entropic term is
determinant of the temperature trend observed for $\Delta G$, which is
qualitatively the same trend observed with the Tersoff model. As
indicated by the convergence trends we discussed in the previous
section for the 0 K enthalpies, a quantitative comparison between the
changes in $\Delta G$ produced by the Tersoff and EDIP potentials
would suffer form the fact that, for the 192-atom cell, the Tersoff
results could be overestimated by as much as a factor of two. Note
that, for silicon, the overall change in $\Delta G$ over the 300 K -
1200 K interval is 18~meV/\AA\ for Tersoff and 9~meV/\AA\ for
EDIP. While speculative, it seems reasonable to expect that the overall
change in $\Delta G$ for the two potentials, over the temperature
range of our calculations, would be in good agreement if we used cells
with larger values of $D$ and $L$.

Regarding the contribution of the entropic term depicted in
Figs.~\ref{G-si} and \ref{G-ge}, two points have to be
emphasized. First of all, the entropy is calculated by taking the
numerical derivative of the free energy, and the numerical data for
this quantity presents some statistical fluctuations. Most certainly,
these fluctuations are bound to be enhanced when one takes numerical
derivatives. This explains some of the oscillations observed in the
entropic term.  Thus the qualitative behavior of the total free-energy
is more representative than the behavior of the isolated entropic term
shown in the figures.  Another point has to do with the behavior of
the entropic term as a function of temperature, in particular for the
computations using the Tersoff potential. The free energy difference
$\Delta G$ is quite small, when compared with the total free energy of
each system, and the same is true for the entropic contribution to
this difference.  These energy differences will be affected by
anharmonic effects, and the discrepancy between the entropic term
computed using the two models (EDIP and Tersoff) above 600 K is
related to the way anharmonic effects are accounted for in each
model. It is clear from Fig.~\ref{bulk} that the results obtained
using the Tersoff potential tend to deviate more from the experimental
data than those obtained using the EDIP, and that the difference
between the two potentials increase with increasing temperature. The
more accentuated increase in the entropic term computed using the
Tersoff potential has to do with its tendency to overestimate the
anharmonic effects, and represents a small amount of energy that is
not related to any changes in the structure of the defect (which would
imply in much bigger changes in the free energy).

In the case of Ge, $\Delta G$ increases with temperature faster than
in Si (compared to either of the Si potentials used), from 2~meV/\AA\
at 300 K to 33~meV/\AA\ at 1200 K. While the enthalpy difference at 0
K marginally favors the SP reconstruction (at variance with the LDA
result in Ref.~\onlinecite{Nunes_3}), the behavior of the free energy
suggests that at room temperature the DP reconstruction would be more
stable than the SP structure. It is important to emphasize the
similarity between our results for Si and Ge, i.e., the free energy
difference $\Delta G$ increases with temperature in both cases (the
increase rate is larger in Ge than in Si), meaning that the entropic
term leads to an enhancement of the thermodynamical stability of the
DP structure with respect to the SP, as the temperature increases.

At this point it is interesting to compare the RS-MC results for Si
with the harmonic-approximation calculations of Valladares {\it et
al.}.~\cite{Vall} In Fig.~\ref{vallad}, we show our RS-MC calculations
using the Tersoff (full line) and EDIP (dashed line) models, and the
Tersoff harmonic-approximation results from Ref.~\onlinecite{Vall}
(dotted line). For a better comparison of the temperature trends in
each calculation, our values for $\Delta G$ for both potentials have
been shifted such that they extrapolate to the same zero-temperature
$\Delta G$ value obtained by Valladares {\it et al.}. The same
qualitative behavior is observed in Fig.~\ref{vallad} for the RS-MC
calculations using both the Tersoff and the EDIP models. In both
cases, $\Delta G$ {\it increases} with temperature, while the opposite
is observed for the harmonic-approximation calculations, which show
$\Delta G$ {\it decreasing} with increasing temperature.  This points
to an important role of anharmonic effects in describing thermal
effects on dislocation cores in semiconductors. 

It should also be pointed out that at low temperatures our
results cannot be directly compared to those in
Ref.~\onlinecite{Vall}, because their calculations take into account
quantum effects which are absent in our approach. While quantum
effects may be relevant at low temperatures, we are mostly concerned
with the anharmonic effects, which become more relevant in the
high-temperature regime. Valladares {\it et al.} point out that the
order of magnitude of the difference in free energies between the SP
and the DP reconstructions is smaller than the thermal energy
($k_{B}T$) over the entire range of temperature in their study. This
would suggest that the two structures would be nearly equally
stable. Our results, however, indicate that at high temperatures the
free energy difference between the SP and the DP cores is of the same
order of the thermal energy. Therefore, considering that the thermal
fluctuations are smaller than the thermal energy itself, our results
suggest that the DP reconstruction should be dominant in the high
temperature regime. In other words, while thermal fluctuations may
account for the creation of local defects such as kinks, it is
unlikely that these fluctuations may cause substantial portions of the
dislocation to switch from the DP to the SP reconstruction (barring a
possible dependence of the core energies on the stress acting on the
dislocation,~\cite{Blase} an issue that has not yet been investigated
in Si and Ge).

\section{Conclusions}              
In this work, we studied the differences in free energies and
vibrational entropies between the SP and DP core reconstructions of
the 90$^{\circ}$ partial dislocation as a function of temperature in
silicon and germanium, using a Reversible Scaling Method and Monte
Carlo simulations, with Tersoff and EDIP models for the
energetics. This methodology is a fully-classical simulation which
includes all anharmonic vibrational effects. Our results indicate that
the difference in the free energies (and in the vibrational-entropy
contribution to this quantity) between the two core structures
increases with temperature in both materials. In the case of Si, this
behavior occurs for both potentials used in the calculations. This is
in contrast with the free-energy calculations by Valladares and
co-workers~\cite{Vall} which incorporates vibrational-entropy effects
only at the harmonic-approximation level. Our calculations indicate
that the DP reconstruction, which is lower in energy at 0 K, becomes
even more stable with respect to the SP structure in the high
temperature regime. Moreover, our results also suggest that the
anharmonic effects may play an important role in the description of
the thermal behavior of extended defects in semiconductors.
  
\begin{acknowledgments} 

C. R. Miranda and A. Antonelli acknowledge the support from the 
Brazilian funding agencies: FAPESP, CNPq, and FAEP.
R. W. Nunes acknowledges support from the Brazilian agencies: CNPq and
FAPEMIG.

\end{acknowledgments}  

%\begin{thebibliography}{99}

\begin{table}
\centering{
\caption{Calculated enthalpy differences at 0 K between core
reconstructions of the 90$^{\circ}$ partial dislocation, for four
different cell sizes, in meV/\AA. Results are shown for TETB (from
Ref.~17), Tersoff, and EDIP. Cells are defined by their height $D$ and
width $L$ (in \AA), as defined in the text. The number of atoms
($N_{at}$) in each cell is also included. $\Delta{E}_{SP}$ is the
difference in enthalpy between the cells with the same senses and
opposite senses of the SP reconstruction (see text).  $\Delta E_{SP -
DP}$ is the difference in enthalpy between the average of the SP cells
and the DP structure.}
\label{tab1}
\begin{tabular}{lcccccccc}  
& & &\multicolumn{2}{c}{Tersoff} &\multicolumn{2}{c}{EDIP} &\multicolumn{2}{c}{TBTE} \\
$~N_{at}$ &$L$ &$D$ &$E_{SP-DP}$ &$\Delta E_{SP}$ &$E_{SP-DP}$ &$\Delta E_{SP}$ &$E_{SP-DP}$ &$\Delta E_{SP}$ \\
\hline
~192 &13.3 &18.8 &~13 &24 &-35 &~6 &~62 &39 \\
~240 &16.6 &18.8 &~~9 &16 &-39 &~4 &~55 &26 \\
~576 &26.6 &28.2 &~~7 &~6 &-39 &~2 &~55 &~9 \\
1920 &53.2 &47.0 &~~6 &~2 &-40 &~0 &~55 &~4 
\end{tabular}
}
\end{table}

\begin{figure}
\caption{Atomic structure of the 90$^{\circ}$ partial dislocation
viewed from above the $\{111\}$ slip plane. (a) symmetrically
reconstructed core; (b) SP reconstruction; and (c) DP
reconstruction.
\label{cores}}
\end{figure} 
          
\begin{figure}
\caption{Gibbs free-energy for the bulk of silicon. Full and dashed
lines show our RS-MC calculations, including anharmonic effects, for
the EDIP and Tersoff potentials, respectively. The dotted line shows the
Tersoff-potential harmonic-approximation results from Ref.~28,
and the squares are the experimental results from Ref.~29.
\label{bulk}}
\end{figure} 
          
\begin{figure}
\caption{Gibbs free-energy difference per unit length $\Delta G$,
between the SP and the DP geometries of the 90$^\circ$ partial
dislocation in silicon, with different samplings of the phonon modes
along the dislocation direction. For the 192-atom calculations, the
results for the SP cores with same reconstruction senses, with
opposite reconstruction senses, and the average between the two (as
explained in the text) are shown by the squares, circles, and
triangles, respectively. For the 384-atom cell, same reconstruction
senses, opposite reconstruction senses, and the average are
represented by the solid line, the dashed line, and the dotted line,
respectively.
\label{tamanho}}
\end{figure} 

\begin{figure}
\caption{Gibbs free-energy difference per unit length ($\Delta G$) and
the entropic contribution $-T\Delta S$, between the SP and the DP
geometries in silicon, in $\rm{meV/\AA}$, using the Tersoff
potential (full line for $\Delta G$ and squares for $-T\Delta S$), and
the EDIP potential (dashed line for $\Delta G$ and circles for
$-T\Delta S$).
\label{G-si}}
\end{figure} 

\begin{figure}
\caption{Gibbs free-energy difference per unit length $\Delta G$, and
the entropic contribution $-T\Delta S$, in $\rm{meV/\AA}$, between the
SP and the DP geometries in germanium, using the Tersoff
potential (full line for $\Delta G$ and squares for $-T\Delta S$).
\label{G-ge}}
\end{figure} 

\begin{figure}
\caption{Difference in free energy between SP and DP cores in silicon for
RS-MC calculations using the Tersoff (full line) and EDIP (dashed
line) potentials, compared with the harmonic-approximation results
from Ref.~10. RS-MC energies at 0 K were shifted, for better
comparison (see text).
\label{vallad}}
\end{figure} 

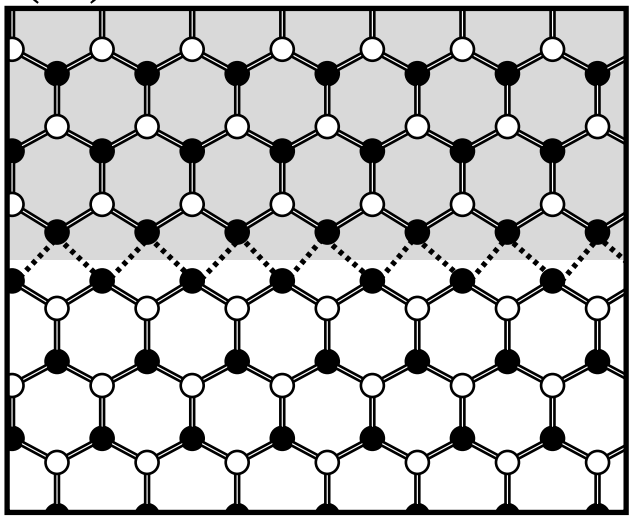
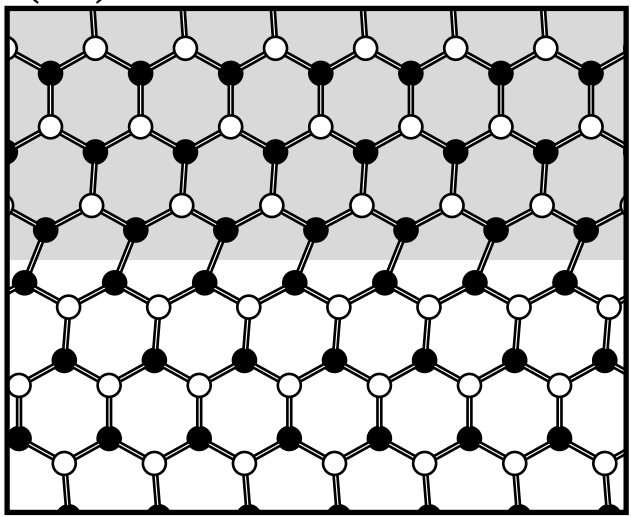
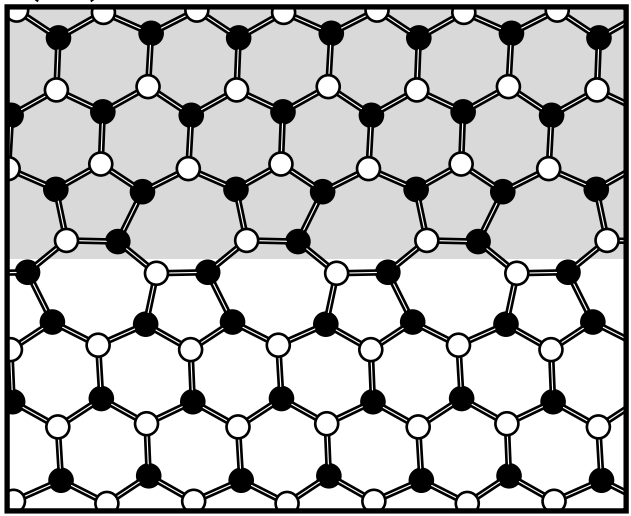
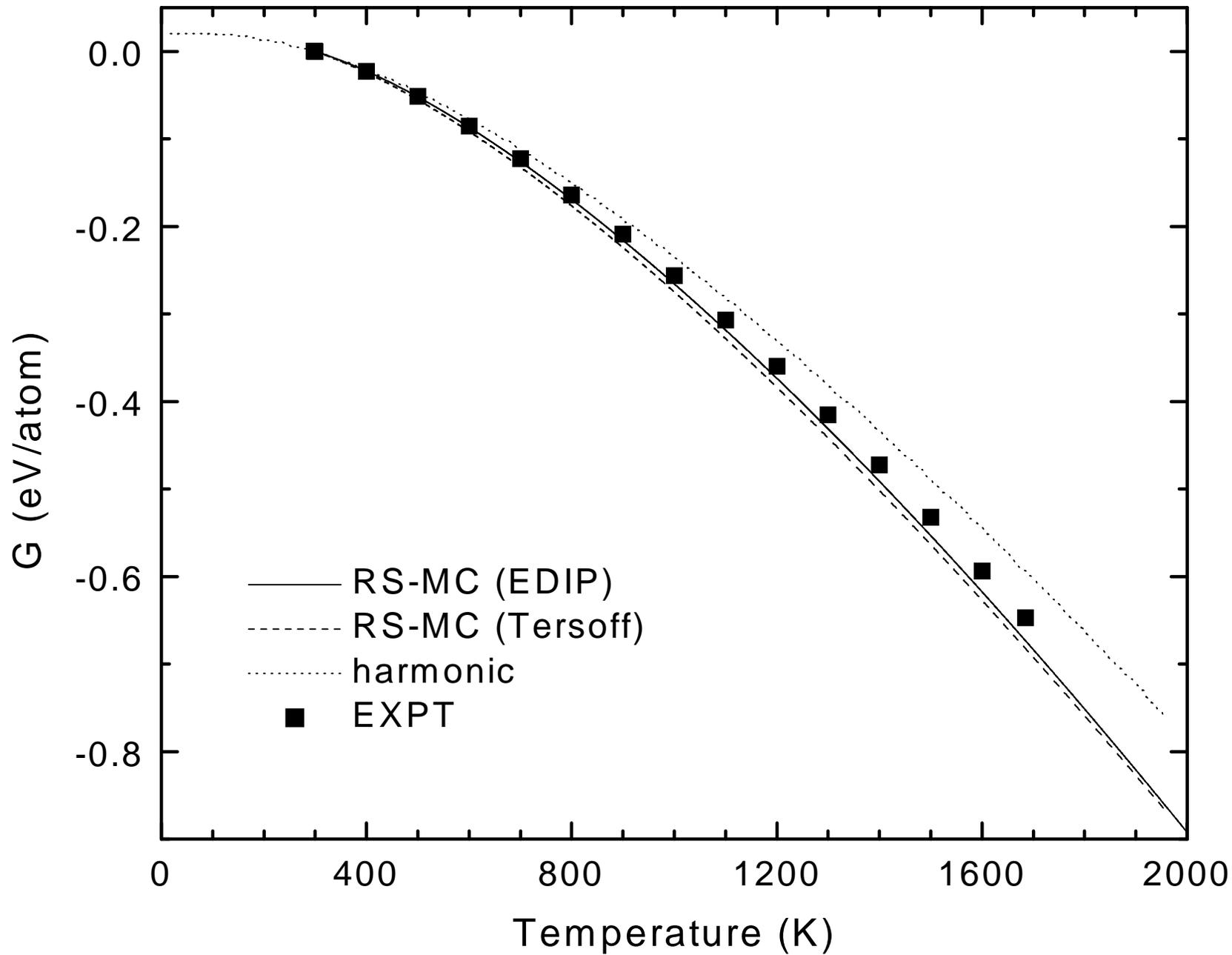
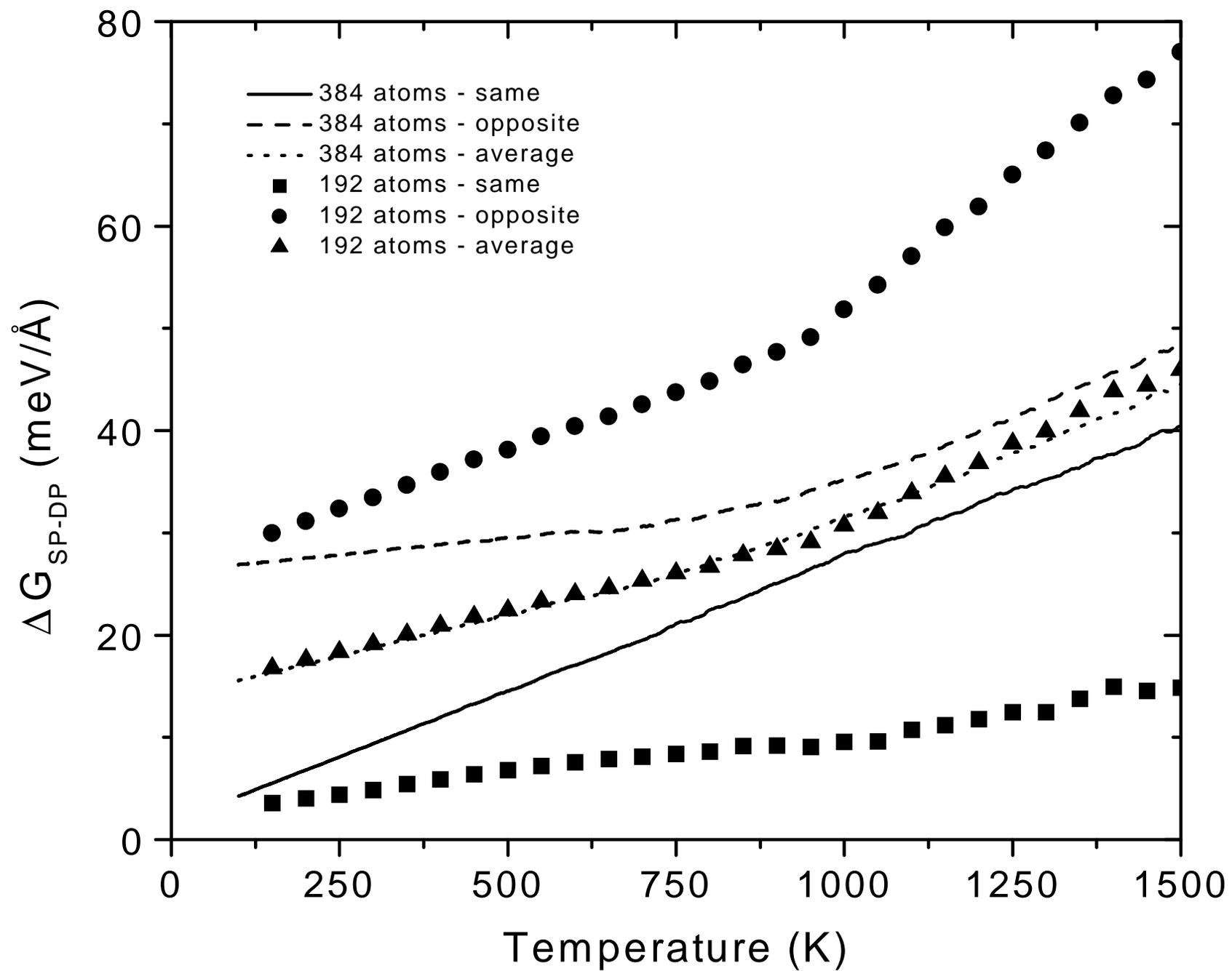
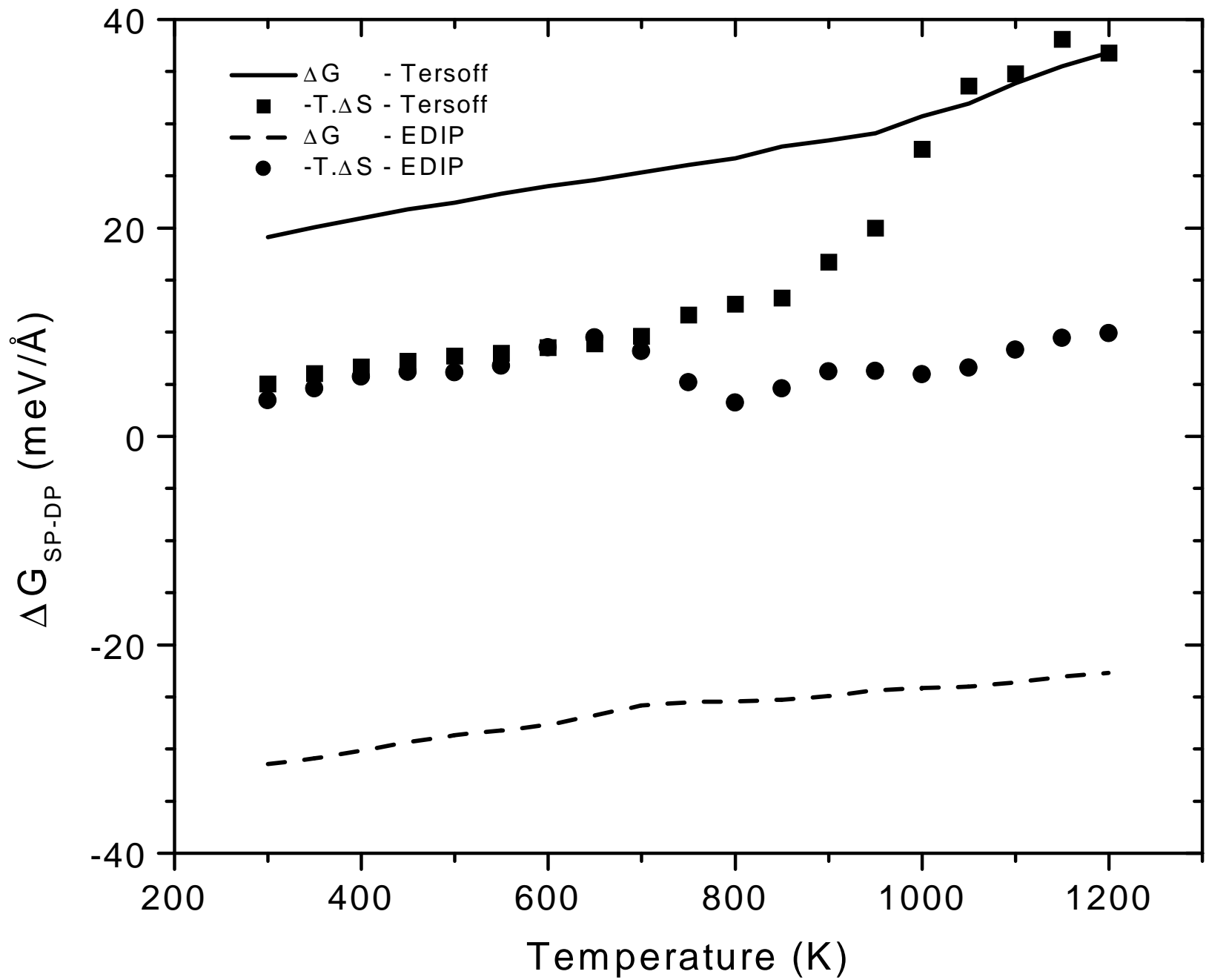
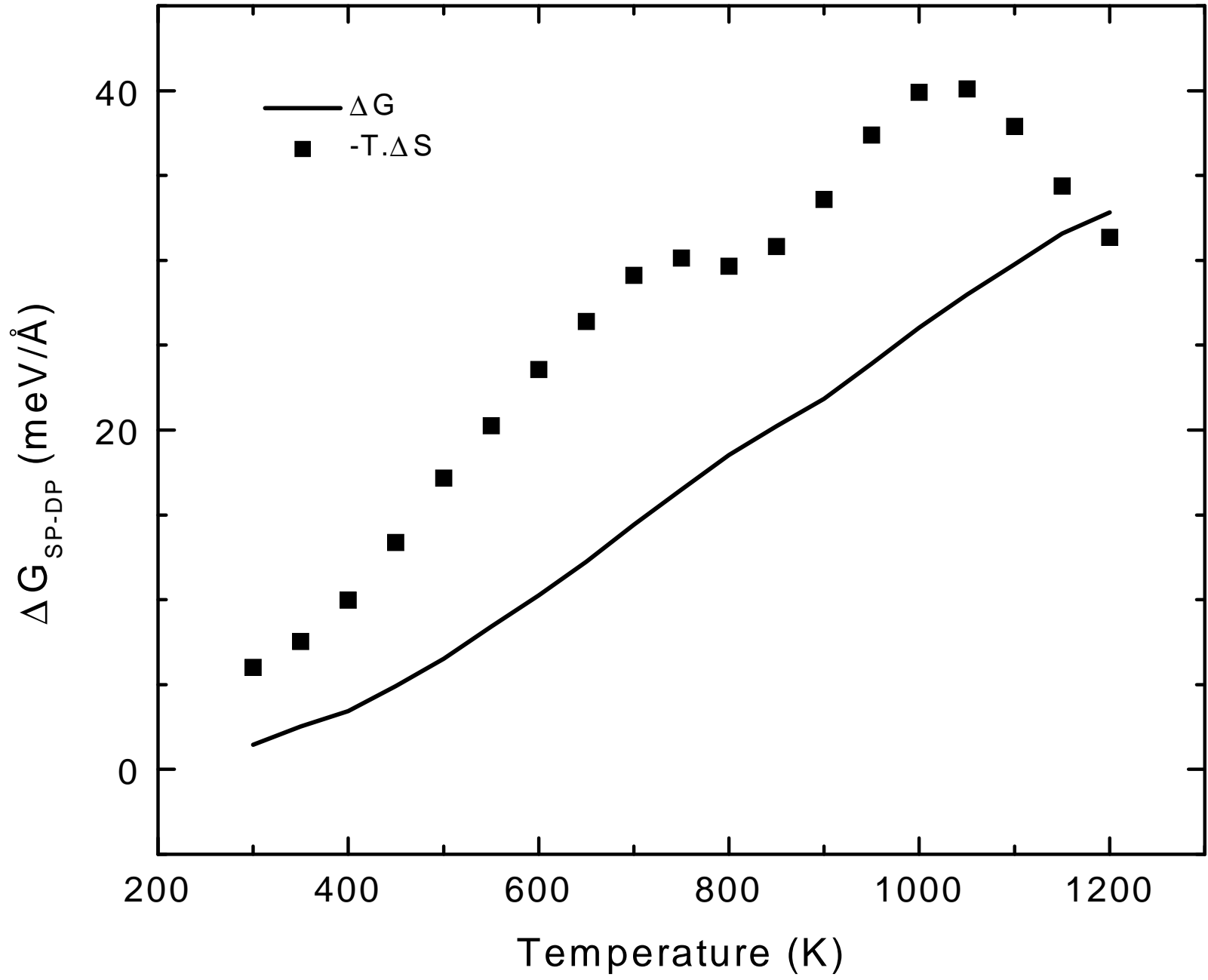
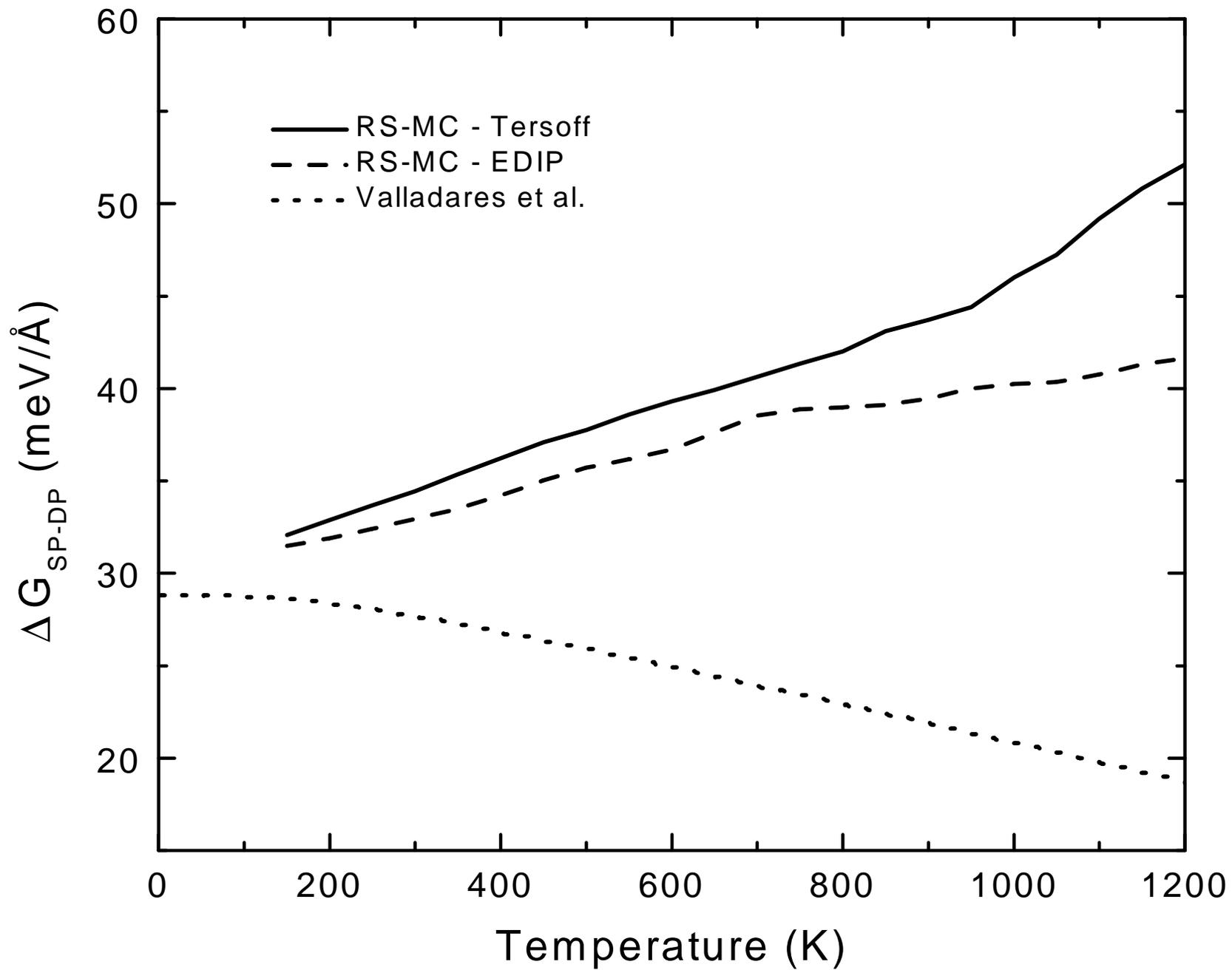
\end{document}